\documentclass[english,twoside,a4paper,10pt]{article}

\usepackage[latin1]{inputenc}
\usepackage[T1]{fontenc}
\usepackage{amsmath}
\usepackage{amsfonts}
\usepackage{graphicx}
\usepackage{subfigure}
\usepackage{a4wide}
\usepackage{amssymb}
\usepackage{fancyhdr}
\usepackage{mathrsfs}

\begin{document}

\title{\bf Topological Static Spherically Symmetric vacuum Solutions in $\mathcal F(R,G)$ Gravity }
\author{ 
R. Myrzakulov$^1$\footnote{Email: rmyrzakulov@gmail.com; rmyrzakulov@csufresno.edu}\,,
L. Sebastiani$^{1}$\footnote{E-mail address:l.sebastiani@science.unitn.it
}\, and S. Zerbini$^2$\footnote{E-mail address:zerbini@science.unitn.it}\\
\\
\begin{small}
$^1$ Eurasian International Center for Theoretical Physics and  Department of General
\end{small}\\
\begin{small} 
Theoretical Physics, Eurasian National University, Astana 010008, Kazakhstan
\end{small}\\
\begin{small}
$^2$ Dipartimento di Fisica, Universit\`a di Trento, Italy and 
\end{small}\\
\begin{small}
 Gruppo Collegato di Trento, Istituto Nazionale di Fisica Nucleare, Sezione di Padova, Italy
\end{small}\\
}

\date{}

\maketitle


\begin{abstract}
The  Lagrangian derivation of the Equations of Motion for topological static spherically symmetric metrics in $\mathcal F (R,G)$-modified gravity is presented and the related solutions are discussed. In particular, a new topological solution for the model $\mathcal F (R,G)=R+\sqrt{G}$ is found.
The black hole solutions and the First Law of thermodynamic are analyzed. 
Furthermore, the coupling with electromagnetic field is also considered and a Maxwell solution is derived.
\end{abstract}



\section{Introduction}

Recent observational data imply -against any previous belief- that the current expansion of the universe
is accelerating~\cite{SN1,WMAP0,WMAP}. Since this discovery, the so called Dark Energy issue has become the ``Mystery
of the Millennium'' \cite{Padmanabhan:2006ag} and today dark
energy is probably one of the most ambitious and tantalizing field of
research because of its implications in fundamental physics. There exist several descriptions of the acceleration of the universe.  Among them, the simplest one is the introduction of small positive 
Cosmological Constant in the framework of General Relativity (GR), the so called $\Lambda$CDM model. 
A generalization of this simple modification of GR consists in considering  modified gravitational theories, 
 where some combination of curvature invariants (the Riemann tensor, the Weyl tensor, the Ricci tensor and so on) replaces, or is added, to the classical Hilbert-Einstein action of GR.
The simplest class of modified theories is $F(R)$-modified gravity, in which the action is described by a function $F(R)$ of the Ricci scalar $R$
(for a review see Refs.~\cite{review6,review7}), but other modifications are in principle allowed.
For example, an interesting class of modified gravity models which may easily produce
the acceleration epoch is string-inspired modified Gauss Bonnet
gravity, the so-called
$F(G)$-gravity,
where $F(G)$ is a function of the four dimensional Gauss Bonnet invariant
$G$. This class of modified gravity models may easily produce the late-time acceleration 
and is an interesting alternative to standard cosmology~\cite{Nojiri:2005jg, F(G)-gravity}.

If some modified theory lies behind our universe, it is of crucial interest to extend the proprieties and the laws of General Relativity to its framework. In this paper, we focus our interest on the topological
static spherically symmetric (SSS) solutions
in modified gravity $\mathcal F(R, G)$-models, that is, the modification of gravity is given by a function of the both, the Ricci scalar and the Gauss-Bonnet invariant.
Typically the modified gravity models admit the de Sitter (dS) space as a solution, but
the number of exact non-trivial (i.e., different to Schwarzshild-de Sitter one) SSS solutions so far known in modified gravity is extremely small, due to the complexity of the field equations.

The SSS solutions in $F(R)$-gravity  have been investigated in several papers (see Refs. \cite{Multamaki,capo,bezerra, saffari, capo2}).  For the specific model $F(R)=R^{1+\delta}$, $\delta$ being a fixed parameter, a class of 
exact SSS solutions has been presented in \cite{CB}. 
For SSS solutions in Weyl gravity see also Refs. \cite{Deser, Pope}.
In Ref. \cite{SSSsolutions} 
a suitable derivation of the $F(R)$-equations of motion based on the lagrangian multilpleyers has been presented and it has been shown how this approach permits to find a large class of solutions. 
Here, we would like to extend this method to the topological case of SSS solutions in $\mathcal F (R,G)$-modified gravity. 
In Section {\bf 2} we present our Lagrangian derivation of the Equations of Motion (EOM) and
in Secs. {\bf 3}-{\bf 4} we will see how this approach permits to find explicit solutions via reconstruction method. In particular, an exact solution with non constant Gauss-Bonnet is found and a large family of solutions with constant Gauss-Bonnet is discussed. 

The SSS solutions may describe the black holes (BH) and in Section {\bf 5} we will evaluate the BH entropy in $\mathcal F(R,G)$-gravity via Wald method \cite{Wald}. Furthermore,
it seems very difficult to address the analogue of the Misner-Sharp BH mass of GR to modified gravity, where the definition of a sensible mass parameter is a debated question. In Section {\bf 6},  by following
the proposal of Ref. \cite{SSSEnergy} for $F(R)$-gravity, we will try to
identified the mass with a quantity proportional to the constant of integration, which appears in the explicit solutions, making use of the First Law of black hole thermodynamics and evaluating independently the entropy via Wald method and the Hawking temperature via quantum mechanical methods in curved space-times \cite{HT}.
An attempt to derive the First Law from the Equations of Motion of $\mathcal F (R,G)$-gravity is done.

In Section {\bf 7} we provide the formalism for topological SSS solutions in the presence of Maxwell field and an exact solution for a specific Gauss-Bonnet modified gravity model is found. 
Conclusions are given in Section {\bf 8}.

We use units of $k_{B}=c=\hbar=1$ and denote the gravitational constant
$\kappa^2=8\pi G_N\equiv8\pi/M_{Pl}^2$ with the Planck mass of
$M_{PL}=G^{-1/2}_N=1.2\times 10^{19}\text{GeV}$.

\section{Lagrangian approach for topological static spherically symmetric vacuum solutions }

In this Section we present a suitable derivation of the  $\mathcal F (R,G)$-equations of motion based on lagrangian multipleyers, which permits to deal with
an ordinary differential equation system.
The action of modified $\mathcal F(R, G)$-theories reads (in vacuum)
\begin{equation}
I=\frac{1}{2\kappa^2}\int_{\mathcal{M}} d^4 x\sqrt{-g}\mathcal{F}(R,G)\,,\label{action} 
\end{equation}
where $g$ is the determinant of metric tensor, $g_{\mu\nu}$, $\mathcal{M}$ is the space-time manifold and $\mathcal F(R,G)$ is a generic function of the Ricci scalar $R$ and the Gauss Bonnet four-dimensional invariant $G$,  
\begin{equation}
G=R^{2}-4R_{\mu\nu}R^{\mu\nu}+R_{\mu\nu\xi\sigma}R^{\mu\nu\xi\sigma}\,.\label{GaussBonnet}
\end{equation}
The Gauss Bonnet is a combination of the Riemann Tensor $R_{\mu\nu\xi\sigma}$, the Ricci Tensor $R_{\mu\nu}=R^{\rho}_{\mu\rho\nu}$ and its trace $R=g^{\alpha\beta}R_{\alpha\beta}$, namely the Ricci scalar. 

We look for static, (pseudo-)spherically symmetric (SSS) solutions
with various topologies, and write the metric element as
\begin{equation}
ds^2=-e^{2\alpha(r)}B(r)dt^2+\frac{d r^2}{B(r)}+r^2\,\left(\frac{d\rho^2}{1-k\rho^2}+\rho^2 d\phi^2\right)\label{metric0}\,,
\end{equation}
where $\alpha(r)$ and $B(r)$ are functions of the radius $r$, and the manifold will be either a sphere $S_2$, a torus $T_2$ or a compact hyperbolic manifold $Y_2$, according to whether $k=1,0,-1$, respectively. With this Ansatz, the scalar curvature and the Gauss Bonnet read
\begin{eqnarray}
R  &=&
-3\,\left[{\frac{d}{dr}}B\left(r\right)\right]{\frac{d}{dr}}
\alpha\left(r\right)-2\,B\left(r\right)\left[{\frac{d}{dr}}
\alpha\left(r\right)\right]^{2}-{\frac{d^{2}}{d{r}^{2}}}
B\left(r\right)-2\,B\left(r\right){\frac{d^{2}}{d{r}^{2}}}\alpha\left(r\right)\nonumber\\
&&-\frac{4}{r}\,\frac{d}{dr}B\left(r\right)
-4\,\frac{B\left(r\right)}{r}\frac{d}{dr}\alpha\left(r\right)-2\,{\frac{B\left(r\right)}{{r}^{2}}}
+\frac{2k}{{r}^{2}}\,,\label{R}
\end{eqnarray}
\begin{eqnarray}
G &=&\frac{4}{r^2}\left[\left(\frac{d\alpha(r)}{dr}\right)\left(\frac{d B(r)}{dr}\right)(5B(r)-3k)+\left(\frac{d B(r)}{dr}\right)^2+\left(\frac{d^2 B(r)}{dr^2}\right)(B(r)-k)\right.\nonumber\\ \nonumber\\
&&\left.+2(B(r)-k)B(r)\left(\left(\frac{d\alpha(r)}{dr}\right)^2 + \frac{d^2\alpha(r)}{dr^2}\right)\right]\,.\label{G} 
\end{eqnarray}
By plugging this expressions into the action (\ref{action}), one obtains a higher derivative Lagrangian theory. 
In order to work with a first derivatives Lagrangian system, we may use the  method of Lagrangian multipliers 
(for the case of FRW space-time see  Refs.~\cite{Monica,Vilenkin,Capozziello}). This method  
permits to consider as independent Lagrangian coordinates 
the scalar curvature $R$, the Gauss Bonnet invariant $G$ and the quantities $\alpha(r)$ and $B(r)$, appearing in the topological spherically static symmetric Ansatz. 

By introducing the Lagrangian multipliers $\lambda$ and $\mu$ and using Eqs.~(\ref{R})-(\ref{G}), the  action may be written as 
\begin{eqnarray}
I &\equiv& \frac{1}{2\kappa^2}\int dt\int d{ r}\left(\mathrm{e}^{\alpha(r)}r^2 \right)\left\{ \mathcal{F}(R,G)-\lambda \left[R+3\,\left({\frac{d}{dr}}B\left(r\right)\right){\frac{d}{dr}}
\alpha\left(r\right)\right.\right.\nonumber\\ \nonumber\\
&& +2\,B\left(r\right)\left({\frac{d}{dr}}
\alpha\left(r\right)\right)^{2}+{\frac{d^{2}}{d{r}^{2}}}
B\left(r\right)+2\,B\left(r\right){\frac{d^{2}}{d{r}^{2}}}\alpha\left(r\right)+\frac{4}{r}\frac{d}{dr}B(r)\nonumber\\ \nonumber\\ 
&&+4\,{\frac{B\left(r\right)}{r}}{\frac{d}{dr}}\alpha\left(r\right)+2\,{\frac{B\left(r\right)}{{r}^{2}}}
\left.-\frac{2k}{{r}^{2}}\right]\nonumber\\ \nonumber\\
&&\hspace{-15mm}-\mu\left[G- 
\frac{4}{r^2}\left[\left(\frac{d\alpha(r)}{dr}\right)\left(\frac{d B(r)}{dr}\right)(5B(r)-3k)+\left(\frac{d B(r)}{dr}\right)^2+\left(\frac{d^2 B(r)}{dr^2}\right)(B(r)-k)\right.\right.\nonumber\\ \nonumber\\
&&\left.\left.+2(B(r)-k)B(r)\left(\left(\frac{d\alpha(r)}{dr}\right)^2 + \frac{d^2\alpha(r)}{dr^2}\right)\right]
\right\}\,.
\end{eqnarray}
Making the variation with respect to $R$ and $G$, one gets
\begin{equation}
\lambda=\frac{\partial}{\partial R}\mathcal{F}(R,G)\,,
\end{equation}
\begin{equation}
\mu=\frac{\partial}{\partial G}\mathcal{F}_G(R,G)\,.
\end{equation}
Thus, by substituting this values and by making an integration by part, the total Lagrangian $\mathscr L$ of the system takes the form\\
\phantom{line}\\
\begin{eqnarray}
&&\hspace{-15mm}\mathscr L(\alpha, d\alpha/dr, B, d B/dr, R, d R/dr, G, d G/dr)=\mathrm{e}^{\alpha(r)}\left\{r^2\left(\mathcal{F}-\mathcal{F}'_R R-\mathcal{F}'_G G\right)\phantom{\frac{1}{1}}\right.
\nonumber\\ \nonumber\\
&&\hspace{15mm}+2\mathcal{F}'_R\left(k-r\frac{d B(r)}{dr}- B(r)\right)
+\mathcal{F}''_{RR}\frac{d R}{d r}r^2\left(\frac{d B(r)}{d r}+2B(r)\frac{d \alpha(r)}{dr}\right)\nonumber\\ \nonumber\\
&&\left.\hspace{15mm}-\mathcal{F}_{GG}''\frac{d G}{dr}\left(4\frac{d B(r)}{dr}+8B(r)\frac{d\alpha(r)}{de}\right)(B(r)-k)\right\}\,.
\end{eqnarray}
\phantom{line}\\
Now, $\mathcal{F}(R,G)$ has been replaced with $\mathcal{F}$ and we have used the following expressions:
\begin{eqnarray}
{\mathcal{F}}'_{R} \equiv
\frac{\partial \mathcal{F}}{\partial R}\,, \quad
{\mathcal{F}}'_{G} \equiv
\frac{\partial \mathcal{F}}{\partial G}\,, 
\label{convention}
\end{eqnarray}
and so on.
The lagrangian depends on the first derivatives of the variables at most.
It is also easy to see that, if $\mathcal{F}'_G=\mathrm{Const}$, i.e. the Gauss Bonnet simply is an additive term, its contribute vanishes.
Making the variation with respect to $\alpha(r)$ and with respect to $B(r)$, one finally gets the Equations of Motion:
\phantom{line}
\begin{eqnarray}
&&\hspace{-5mm}\mathrm{e}^{\alpha(r)}\left\{r^2(\mathcal{F}-\mathcal{F}'_R R-\mathcal{F}'_G G)+2\mathcal{F}'_R\left[k-r\left(\frac{d B(r)}{dr}\right)-B(r)\right]-\frac{d\mathcal{F}'_R}{dr}\left[r^2 \left(\frac{d B(r)}{dr}\right)+4rB(r)\right]\right. \nonumber\\ \nonumber\\
&&\hspace{-0mm}\left.-2r^2B(r)\frac{d^2\mathcal{F}'_R}{dr^2}+4(3B(r)-k)\left(\frac{d B(r)}{dr}\right)\frac{d\mathcal{F}'_G}{dr}
+8B(r)(B(r)-k)\frac{d^2\mathcal{F}'_G}{dr^2}\right\}=0\,,\label{EOM1SphRG}\\ \nonumber\\
&&\hspace{-5mm}\mathrm{e}^{\alpha(r)}\left\{\frac{d\alpha(r)}{dr}\left(2r\mathcal{F}'_R+r^2\frac{d\mathcal{F}'_R}{dr}-4(3B(r)-k)\frac{d\mathcal{F}'_G}{dr}\right)-r^2\frac{d^2\mathcal{F}'_R}{dr^2}+4(B(r)-k)\frac{d^2\mathcal{F}'_G}{dr^2}\right\}=0\label{EOM2SphRG}. 
\end{eqnarray}
The above equations with Eqs.~(\ref{R})-(\ref{G}) form a system of four ordinary differential equations in the four unknown quantities $\alpha(r)$, $B(r)$, $R=R(r)$ and $G=G(r)$. 
In principle, given the model $\mathcal F(R,G)$ as a function of $r$, our expressions permit to derive both $\alpha(r)$ and $B(r)$, that is the explicit form of the metric. On the other hand, it is also possible to try to reconstruct the models by starting from the solutions. For example,
fixing the form of $\alpha(r)$ one may reconstruct the model by using Eq. (\ref{EOM2SphRG}) and therefore $B(r)$ from Eq. (\ref{EOM1SphRG}). 
In general, the Lagrangian $F(R)$ one eventually finds is not unique since one has to
infer its form starting from the value it assumes on the solutions (see Section {\bf 4}).
We observe that by explicitly written $R$ and $G$ as functions of $B(r)$ and $\alpha(r)$, we reduce the system to the usual high order differential equations system of $\mathcal F(R,G)$-gravity. Since in this paper we would like to explore (topological) SSS solutions in Gauss Bonnet gravity, in what follows we will always consider the cases of non trivial contribution of $G$ into the action.

\section{Solutions with constant $\alpha(r)$}

In order to see how the procedure of reconstruction works, 
we first divide Eq.~(\ref{EOM1SphRG}) with respect to $\mathrm{e}^{\alpha(r)} r^2$ and then
perform the derivative with respect to $r$, obtaining\\
\phantom{line}
\begin{eqnarray}
&&-\left(\frac{d}{d r}\mathcal F_R'\right) R-\left(\frac{d}{d r}\mathcal F_G'\right) G
+2\mathcal F'_R\left(\frac{3B(r)}{r^3}-\frac{2k}{r^3}-\frac{1}{r}\frac{d^2B(r)}{d r^2}\right)
\nonumber\\ \nonumber\\
&&+\left(\frac{d}{d r}\mathcal F_R'\right)\left(\frac{2B(r)}{r^2}+\frac{2k}{r^2}-\frac{6}{r}\frac{dB(r)}{d r}-\frac{d^2 B(r)}{d r^2}\right)
-\left(\frac{d^2}{d r^2}\mathcal F_R'\right)\left(3\frac{d B(r)}{d r}+\frac{4 B(r)}{r}\right)
\nonumber\\ \nonumber\\
&&-2\left(\frac{d^3}{d r^3}\mathcal F_R'\right)B(r) 
+4\left(\frac{d}{d r}\mathcal F_G'\right) \left[\left(\frac{3}{r^2}\frac{dB(r)}{d r}
-\frac{6 B(r)}{r^3}+\frac{2k}{r^3}
\right)\frac{d B(r)}{d r}+
\left(\frac{3B(r)}{r^2}-\frac{k}{r^2}\right)\frac{d^2B(r)}{d r^2}
\right]
\nonumber\\ \nonumber\\
&&+\left(\frac{d^2}{d r^2}\mathcal F_G'\right)
\left(\frac{28 B(r)}{r^2}\frac{dB(r)}{d r}-\frac{12 k }{r^2}\frac{dB(r)}{d r}
-\frac{16 B(r)^2}{r^3}+\frac{16k B(r)}{r^3}\right)\nonumber\\ \nonumber\\&&
+8\left(\frac{d^3}{d r^3}\mathcal F_G'\right)B(r)\left(\frac{B(r)}{r^2}-\frac{k}{r^2}\right)=0\,.
\label{new}
\end{eqnarray}
\phantom{line}\\
The utility of this equation is due to the fact that it depends on $\mathcal F_R$ and $\mathcal F_G$ only,
which may be derived from Eq. (\ref{EOM2SphRG}) ones $\alpha(r)$ is fixed.
Let us consider the important case of $\alpha(r)=0$ (it is equivalent to $\alpha(r)=\mathrm{Const}$) into the metric (\ref{metric0}), namely
\begin{equation}
ds^2=-B(r)dt^2+\frac{d r^2}{B(r)}+r^2\,\left(\frac{d\rho^2}{1-k\rho^2}+\rho^2 d\phi^2\right)\label{metric00}\,.
\end{equation}
The equations (\ref{R})-(\ref{G}) lead to:
\begin{equation}
R  =
-{\frac{d^{2}}{d{r}^{2}}}
B\left(r\right)
-\frac{4}{r}\frac{d B(r)}{d r}
-2\,{\frac{B\left(r\right)}{{r}^{2}}}
+\frac{2k}{{r}^{2}}\,,\label{R2}
\end{equation}
\begin{equation}
G=\frac{4}{r^2}\left(\frac{d B(r)}{d r}\right)^2+\frac{4}{r^2}\frac{d^2B(r)}{d r^2}(B(r)-k)\,.\label{G2}
\end{equation}
>From Eq.(\ref{EOM2SphRG}) one has
\begin{equation}
-r^2\frac{d^2}{dr^2}\mathcal F_R'+4(B(r)-k)\frac{d^2}{d r^2}\mathcal F_G'=0\,. 
\end{equation}
A simple choice is to investigate the models with $d^2 \mathcal F_R/d r^2=0$ and $d^2\mathcal F'_G/d r^2=0$, namely
\begin{equation}
\mathcal F'_R=ar+b\,,\quad  \mathcal F'_G= c r\,.\label{Fprime}
\end{equation}
Here, $a$, $b$ and $c$ are integration constants (the pure constant in $\mathcal F'_G$ does not give contribution to (\ref{new}) and will be analyzed in the next Section).
Eq. (\ref{new}) leads to\\
\phantom{line}\\
\begin{eqnarray}
&&-\frac{d^2 B(r)}{dr^2}\left(a+\frac{b}{r}\right)+\frac{2 a}{r^2}\left(2B(r)-k\right)+\frac{2b}{r^3}(B(r)-k)-\frac{a}{r}\frac{d B(r)}{d r}\nonumber\\
&&+c\left(\frac{4}{r^2}\left(\frac{d B(r)}{d r}\right)^2
+\frac{4 B(r)}{r^2}\frac{d^2 B(r)}{d r^2}-\frac{4(3B(r)-k)}{r^3}\frac{d B(r)}{d r}
\right)=0\,.\label{genericequation} 
\end{eqnarray}
\phantom{line}\\
Unfortunately, this equation can not be solved for general values of the parameters. Otherwise, it may be used to find some specific solution.
For example, if $a=0$ and $b=1$, a solution is
\begin{equation}
B(r)=-k+C\,r\,,\label{SSSprime}
\end{equation}
with $c=1/(4C)$. From Eqs. (\ref{R2})-(\ref{G2}) one has
\begin{equation}
R=\frac{4 k-6 C r}{r^2}\,,
\end{equation}
\begin{equation}
G=\frac{4 C^2}{r^2}\,.
\end{equation}
If we use the fact that $r=2C/\sqrt{G}$, we can finally solve Eq.~(\ref{EOM1SphRG}) with respect to $\mathcal F(R,G)$.
As a result, we find that the model\\
\phantom{line} 
\begin{equation}
\mathcal F(R,G)=R+\sqrt{G}\,,\label{tata}
\end{equation}
\phantom{line}\\
generates the (topological) SSS solution (\ref{metric00}) with $B(r)$ given by (\ref{SSSprime}).

\section{Solutions with constant Gauss Bonnet invariant}

We will check for topological SSS solutions wich lead to a constant value of the Gauss Bonnet.
We start by considering the case $\alpha(r)=0$ of the metric (\ref{metric00}).
By putting $G=G_0$, 
where $G_0=\text{Const}$, in Eq. (\ref{G}), one derives the form of $B(r)$,
\begin{equation} 
B(r)=k\pm\frac{\sqrt{G_0 r^4+C_1+C_2\,r}}{2\sqrt{6}}\,,\label{Uno}
\end{equation}
where $C_1$ and $C_2$ are integration constants. 
Now, we have two very simple cases. The first one corresponds to $\mathcal F'_R=0$, it means that the model $\mathcal F(R,G)=F(G)$ depends on $G$ only and we deal with a pure Gauss Bonnet theory. Thus,  Eq. (\ref{EOM2SphRG}) trivially is satisfied and Eq. (\ref{EOM1SphRG}) reads
\begin{equation}
F(G_0)-G_0 F'_G(G_0)=0\,.\label{**}
\end{equation}
Of course, there are infinity functions $F(G)$ which reduces (on shell) to Eq. (\ref{**})  for $G=G_0$.
For example, one finds that the specific model
\begin{equation}
F(G)=\gamma\mathrm{e}^{G/G_0}\,,
\end{equation}
where $\gamma$ is a dimensional constant,
exibits the topological SSS solution (\ref{metric00}) described by $B(r)$ in Eq. (\ref{Uno}), but
also the class of models
\begin{equation}
F(G)=\gamma G^n+\Lambda\,,
\end{equation}
where $\Lambda=\gamma(n-1)G_0^n$, exhibits the same kind of solution.

An other simple possibility is to take constant the Ricci scalar also, $R=R_0$, where $R_0=\text{Const}$. As a consequence of Eq. (\ref{R}), it follows that
$C_1=C_2=0$ in the solution (\ref{Uno}) and $R_0=\sqrt{6 G_0}$. Also in this case, Eq. (\ref{EOM2SphRG}) trivially is satisfied and Eq. (\ref{EOM1SphRG}) reads
\begin{equation}
\mathcal F(R_0,G_0)-R_0\mathcal{F}'_R(R_0,G_0)-G_0 \mathcal F'_G(R_0 G_0)+2\mathcal{F}'_R(R_0,G_0)\sqrt{\frac{3G_0}{8}}=0\,.\label{dScondition}
\end{equation}
It means that in principle every $\mathcal F(R,G)$-model admits the de Sitter solution
\begin{equation}
B(r)=\left(k-\frac{\Lambda
r^2}{3}\right)\,,
\end{equation}
where $\Lambda=\sqrt{3G_0/8}$ is fixed by Eq. (\ref{dScondition}), such that $R_0=4\Lambda$ and $G_0=8\Lambda^2/3$.
For example, the model $\mathcal F(R,G)=R+\gamma G^2$, with $\gamma$ a dimensional constant, exhibits the de Sitter solution with $\Lambda=[9/(32\gamma)]^{1/3}$.
This is a well-known result.\\
\\
Finally, we observe an interesting invariance of the EOM, occurring when $G_0=0$. If the model assumes the form
\begin{equation}
\mathcal{F}(R,G)=G f(R,G)\,,\label{modellino}
\end{equation}
where $f(R,G)$ is a function of $R$ and $G$ again and $\lim_{G\to0} Gf(R,G)=0$, the two Equations of Motion (\ref{EOM1SphRG})-(\ref{EOM2SphRG}) trivially are satisfied. Let us have a look for some examples.

\subsubsection*{Case $\alpha(r)=0$ and $G_0=0$}

If $\alpha(r)=0$, by putting $G_0=0$
in Eq. (\ref{Uno}), we get
\begin{equation}
B(r)=k\pm C_1\sqrt{1+C_2r}\,,\label{Uno2}
\end{equation}
where we have redefined the constants.
The models in the form of (\ref{modellino}) admit this kind of SSS solutions.

\subsubsection*{Case $\alpha(r)\neq 0$ and $G_0=0$}

We simply can also verify that $G_0=0$ provided by
\begin{equation}
B(r)=k\,,\quad k=1\,,
\end{equation}
independently on $\alpha(r)$. In the topological case $k=1$ (the sphere), the models in the form of (\ref{modellino}) admit every SSS solution with $B(r)=1$ and $\alpha(r)$ a general function. This class of solutions can not describe the black holes.

\section{Black hole solutions and Wald entropy}

We may define a black hole event horizon as soon as there is positive solution $r_H$ of
\begin{equation}
B(r_H)=0\,,\quad \left.\frac{d B(r)}{d r}\right|_{r_H}> 0\,. \label{BHconditions}
\end{equation}
The second condition needs to be imposed to preserve the metric signature out of the horizon.
So, for the metric in Eq. (\ref{SSSprime}), it turns out that only in the topological case with $k=1$ (sphere) we can describe a black hole if $C>0$. For metrics in Eqs. (\ref{Uno})-(\ref{Uno2}), we can have black hole solutions by choosing the appropriate sign in $B(r)$ according with the topology. In what follows, it is always understood that we will suppose to deal with SSS solutions which effectively describe the black holes.\\
\\
The entropy associated to the black holes solutions 
can be calculated via Wald's method~\cite{Wald}.
Following Refs.~\cite{Visser:1993nu, FaraoniEntropy}, the explicit calculation of the black hole entropy $S_W$ is provided by the formula
\begin{equation}
S_W = - 2\pi
\oint_{\tiny{\begin{array}{cc}
r=  r_H \\
t = \mbox{const}
\end{array}}}
\left. \left(\frac{\delta \mathscr L}{\delta R_{\mu\nu\xi \sigma}}\right)\right|_{H}\,
e_{\mu \nu} e_{\sigma \xi}\sqrt{h_2}\, d\rho\, d\phi\,.
\label{Wald}
\end{equation}
We use the suffix `$H$' for all quantities evaluated on the horizon. Here, $\mathscr L = \mathscr L(R_{\mu\nu \xi \sigma}, R_{\mu\nu}, R,$ $g_{\mu\nu}, \nabla_\gamma R_{\mu\nu\xi\sigma}...) $ is the Lagrangian density of any general theory of gravity.
The antisymmetric variable
$e_{\mu\nu}=-e_{\nu\mu}$
is the binormal vector to the (bifurcate) horizon.
It is normalized so that $e_{\mu\nu}e^{\mu\nu}=-2$ and, by using the metric Ansatz (\ref{metric0}), it turns out to be 
\begin{equation}
\epsilon_{\mu\nu}=\mathrm{e}^{\alpha(r)}(\delta^0_{\mu}\delta^1_{\nu}-\delta^1_{\mu}\delta^0_{\nu})\,,
\end{equation}
$\delta^i_j$ being the Kronecker delta.
Finally, the induced volume form on the bifurcate surface $r = r_H , t =\mathrm{const}$ is represented by $\sqrt{h_2}\, d\rho\, d\phi$.
The variation of the Lagrangian density with respect to $R_{\mu\nu \xi \sigma}$ is performed as if $R_{\mu\nu \xi \sigma}$ and the metric $g_{\mu\nu}$ were independent. 
Since for the models under consideration $\mathscr L=\mathcal F(R,G)/(2\kappa^2)$, formula (\ref{Wald}) becomes
\begin{eqnarray}
S_W &=& -8\pi \mathcal{A}_H\,
\mathrm{e}^{2\alpha (r_H)}\,\left(\frac{\delta \mathscr L}{\delta R_{0 1 0
1}}\right)\Big\vert_H\nonumber\\
&=&-\frac{8\pi \mathcal{A}_H\mathrm{e}^{2\alpha(r)}}{2\kappa^2}\left(\mathcal F'_R\frac{\delta R}{\delta R_{0 1 0 1}}+\mathcal F'_G\frac{\delta G}{\delta R_{0 1 0 1}}\right)\Big\vert_H\label{waldbis}\,.
\end{eqnarray}
Above,  $\mathcal A_H=V_k r_H^2$, in which $V_1=4\pi$ (the sphere),
$V_0=|\Im\,\tau|$, with $\tau$ the Teichm\"{u}ller
parameter for the torus, and finally
$V_{-1}=4\pi g$, $g>2$, for the compact hyperbolic manifold with genus $g$~\cite{Vanzo}.
Since
\begin{equation}
\frac{\delta R}{\delta R_{\mu \nu \alpha \beta }}=
\frac{1}{2}\left(g^{\alpha \mu}g^{\nu \beta}-g^{\nu \alpha}g^{\mu \beta}  \right)\,,
\end{equation}
\begin{equation}
\frac{\delta G}{\delta R_{\mu\nu\xi\sigma}}=
\left[ 2 R^{\mu\nu\xi\sigma} -2 (g^{\mu \xi} R^{\nu \sigma} + g^{\nu
\sigma} R^{\mu \xi} - g^{\mu \sigma} R^{\nu \xi} - g^{\nu \xi} R^{\mu \sigma})+
(g^{\mu \xi} g^{\nu \sigma} - g^{\mu \sigma} g^{\nu \xi}) R \right]\,,
\label{variation}
\end{equation}
by using the horizon condition $B(r_H)=0$ one obtains:
\begin{equation}
S_W=\frac{\mathcal A_H}{4 G_N}\left[\mathcal F'_R+\mathcal F'_G\left(\frac{4k}{r^2}\right)\right]\Big\vert_H\,.\label{entropy}
\end{equation}
However, the above expression is not manifestly positive. The possible appearance of negative entropy associated with a black hole solution in higher derivative gravity has been discussed in detail in reference \cite{miriam}.  
In General Relativity $\mathcal F(R,G)=R$, $F'_R(R,G)=1 >0$ and one recovers the usual Area Law, namely $S_W=\mathcal A_H/4G_N$.

\section{First Law of thermodynamics}

Evaluating the first equation of motion of $\mathcal F(R,G)$-gravity (\ref{EOM1SphRG}) on the event horizon,
and multiplying both sides of the equation by $V_k/(2\kappa^2)$, we directly obtain
\begin{eqnarray}
\frac{e^{\alpha(r_H)}}{4\pi}\frac{d B(r)}{d r}\Big\vert_H \frac{\partial S_W}{\partial r_H} &=& \nonumber\\ \nonumber\\
&&\hspace{-40mm}\mathrm{e}^{\alpha(r_H)}\left(\frac{k\,\mathcal F_R'(R_H, G_H)}{2G_N}-\frac{R_H\mathcal F_R'(R_H, G_H)+G_H\mathcal F'_G (R_H,G_H)-F(R_H, G_H)}{4G_N}r_H^2\right)\frac{V_k}{4\pi}\,,
\label{EquazionePrincipe}
\end{eqnarray}
where $R_H$ and $G_H$ are the Ricci scalar and the Gauss Bonnet invariant evaluated on the horizon and $S_W$ is the Wald entropy given by Eq.~(\ref{entropy}). Here, we have used the BH horizon condition $B(r_H)=0$.

In 1974 Hawking found an important correlation between the proprieties of black holes and the laws of thermodynamics \cite{HT}. The black holes can irradiate due to quantum effects, that is the so called Hawking radiation.
For static black hole described by the metric~(\ref{metric0}), the Killing/Hawking temperature of the related horizon  reads
\begin{equation}
T_K:=\frac{e^{\alpha(r)}}{4\pi}\frac{d B(r)}{dr}\Big\vert_{H}\,.\label{HT}
\end{equation}
This is a well-known result, and it can be justified in several ways,
for example  making use of standard derivations of Hawking radiation~\cite{VisserHaw},
or by eliminating the conical singularity in the corresponding Euclidean metric,
or either by making use of tunneling methods,
recently introduced in Refs.~\cite{PW,Nadalini}, and discussed in details in several papers.

Thus, in analogy with the case of $F(R)$-gravity treated in Ref. \cite{SSSEnergy}, one may try to derive the
First Law of $\mathcal F(R,G)$-black hole thermodynamics from Eq.~(\ref{EquazionePrincipe}), where the Killing temperature appears in a natural way.  
If the entropy depends on $r_H$ only (it means, when $\mathcal F'_{R,G}\neq\mathrm{Const}$, $R_H$ and $G_H$ do not depend on the integration constants of the solution), we can write
\begin{eqnarray}
T_K \Delta S_W &=&
\nonumber\\ \nonumber\\
&&\hspace{-30mm}\mathrm{e}^{\alpha(r_H)}\left(\frac{k\,\mathcal F_R'(R_H, G_H)}{2G_N}-\frac{R_H\mathcal F_R'(R_H, G_H)+G_H\mathcal F'_G (R_H,G_H)-F(R_H, G_H)}{4G_N}r_H^2\right)\frac{V_k}{4\pi}dr_H\,.
\label{EquazionePrincipe2}
\end{eqnarray}
In this case, the
First Law of $\mathcal F(R,G)$-black hole thermodynamics reads
\begin{equation}
\Delta E_{K}:=T_{K}\Delta S_W\,,\label{differentialform}
\end{equation}
where $\Delta E_{K}$ is the variation of Killing energy $E_K$, and
\begin{eqnarray}
 E_{K}:&=&\nonumber\\ \nonumber\\
&&\hspace{-25mm}\frac{V_k}{4\pi}\int\, \mathrm{e}^{\alpha(r_H)}\left(\frac{k\,\mathcal F_R'(R_H, G_H)}{2G_N}-\frac{R_H\mathcal F_R'(R_H, G_H)+G_H\mathcal F'_G (R_H,G_H)-F(R_H, G_H)}{4G_N}r_H^2\right)dr_H.
\label{BHEnergy0}
\end{eqnarray}
Of course, all this is reasonable only if $r_H$ depends on a unique 
variable parameter, which will be identified with the mass of the black hole.
If $r_H$ and as a consequence also the entropy depends on other variables, 
then other thermodynamics potentials will appear and the expression of energy 
will be different. 

Unlike to the case of pure $F(R)$-gravity \cite{SSSsolutions}, in Gauss Bonnet modified gravity the BH entropy may depend on the integration constant of the solution. It is the case of the model (\ref{tata}), for which
the first law of thermodynamic can not been recovered by starting from the EOM, being Eq. (\ref{BHEnergy0}) not valid. 
However, since in this case the parameter $C=C(r_H)$ which appears in (\ref{SSSprime}) is the solution of $B(r_H)=0$, we may evaluate the variation of entropy as
\begin{equation}
\Delta S_W(r_H, C)=\partial_{r_H} S_W(C,r_H)+\partial_{C} S_W(C,r_H)\frac{d C}{d r_H}\,, 
\end{equation}
and the First Law of thermodynamic holds true as soon as
Eq. (\ref{differentialform}) can be written.
In this example, we will see that the First Law 
permits to
identify the mass with a quantity proportional to the constant of integration, which explicitly appears in the black hole solution.

\subsection*{Examples}

Let us consider the model $\mathcal F(R,G)=R+\sqrt{G}$  of Eq.(\ref{SSSprime}) with the SSS solution $B(r)=-1+C r$ and $G=4C^2/r^2$, which corresponds to the topological case $k=1$ of (\ref{SSSprime}). We remember that for $k=0,-1$ the metric does not describe a black hole. The BH event horizon is defined by 
\begin{equation}
r_H=\frac{1}{C}\,,
\end{equation}
such that $C=1/r_H$. In this case the First Law leads to
\begin{equation}
\Delta E_K:=\frac{d r_H}{G_N}\,,
\end{equation}
and one has
\begin{equation}
E_K=\frac{1}{G_N}\left(\frac{1}{C}\right)>0\,,
\end{equation}
permitting the identification of the energy with the integration constant of the solution.
In this case, the solution is strictly related with the existence of the BH, which could be removed only in the limit $C\rightarrow\infty$. That result is crucial and shows that the solution found is not patological, namely the fact that the mertric changes the signature for $C=0$ is not surpraising, since it means that we are considering a BH with infinitive mass.

In the case of the models with constant Gauss Bonnet solution, we have from Eq. (\ref{entropy}) that the contribute of the Gauss-Bonnet to the entropy is a constant. As a consequence, for the
model $F(G)=\gamma\mathrm{e}^{G/G_0}$ or $F(G)=\gamma G^n+\Lambda$,  $\Lambda=\gamma(n-1)G_0^n$, with BH solution (\ref{Uno}), one obtains $\Delta S_W=0$ and Eq. (\ref{differentialform}) gives us $\Delta E_K=0$.  
Also for the class of models (\ref{modellino}) with solution (\ref{Uno2}), we can use Eq. (\ref{BHEnergy0}) to evaluate the Killing energy and it is easy to see that the result is
$E_K=0$ (in this case $S_W=0$)
and the Firts Law trivially is satisfied.

\section{Maxwell SSS solutions in Gauss-Bonnet modified gravity}

In this Section, we will extend our formalism by considering topological SSS $\mathcal F (R,G)$-solutions in the presence of Maxwell field. We will also provide an exact solution for a specific model of Gauss-Bonnet modified gravity. 
The Maxwell action reads
\begin{equation}
I_{EM}=\frac{1}{4}\int_{\mathcal{M}} d^4 x\sqrt{-g}\,F^{ij}F_{ij}\,,\quad\nabla_k F^{kj}=0\,,
\label{action0em}\end{equation}
where $i,j...$ run from $0$ to $3$, ${\nabla}_{\mu}$ is the covariant derivative
operator associated with the metric $g_{\mu \nu}$ and
$F_{ij}$ is the electromagnetic field strength.
In the case of metric (\ref{metric0}),
due to the symmetry, it is easy to see that the only non vanishing 
component of the electroagnetic field is
\begin{equation}
F_{01}=\frac{e^{\alpha(r)}Q}{r^2}\,,\quad
   F^{01}=-\frac{e^{-\alpha(r)}Q}{r^2}\,,\label{relations}
\end{equation}
$Q$ being the electric charge of electromagnetic field.
The model is now described by the action
\begin{equation}
I=\frac{1}{2\kappa^2}\int_{\mathcal{M}} d^4 x\sqrt{-g}\mathcal{F}(R,G)+\frac{1}{4}\int_{\mathcal{M}} d^4 x\sqrt{-g}\,F^{ij}F_{ij}\,.\label{action2} 
\end{equation}
By using the relations (\ref{relations}) and by following the same lagrangian derivation of Section {\bf 2}, we have that the EOM are equals to (\ref{EOM1SphRG})-(\ref{EOM2SphRG})
with an additive term in (\ref{EOM1SphRG}) due to the electromagnetic field:\\
\phantom{line}
\begin{eqnarray}
&&\hspace{-5mm}\mathrm{e}^{\alpha(r)}\left\{r^2(\mathcal{F}-\mathcal{F}'_R R-\mathcal{F}'_G G)+2\mathcal{F}'_R\left[k-r\left(\frac{d B(r)}{dr}\right)-B(r)\right]-\frac{d\mathcal{F}'_R}{dr}\left[r^2 \left(\frac{d B(r)}{dr}\right)+4rB(r)\right]\right. \nonumber\\ \nonumber\\
&&\hspace{-0mm}\left.-2r^2B(r)\frac{d^2\mathcal{F}'_R}{dr^2}+4(3B(r)-k)\left(\frac{d B(r)}{dr}\right)\frac{d\mathcal{F}'_G}{dr}
+8B(r)(B(r)-k)\frac{d^2\mathcal{F}'_G}{dr^2}-\frac{\tilde Q^2}{2 r^2}\right\}=0\,,\label{EOM1}\\ \nonumber\\
&&\hspace{-5mm}\mathrm{e}^{\alpha(r)}\left\{\frac{d\alpha(r)}{dr}\left(2r\mathcal{F}'_R+r^2\frac{d\mathcal{F}'_R}{dr}-4(3B(r)-k)\frac{d\mathcal{F}'_G}{dr}\right)-r^2\frac{d^2\mathcal{F}'_R}{dr^2}+4(B(r)-k)\frac{d^2\mathcal{F}'_G}{dr^2}\right\}=0\label{EOM2}. 
\end{eqnarray}
\phantom{line}\\
Here, we have defined $\tilde Q^2=2\kappa^2 Q^2$. 

Let us see for an explicit example and consider the pure Gauss-Bonnet gravity model $\mathcal F(R,G)=F(G)$, which depends on $G$ only, with the metric Ansatz $\alpha (r)=0$ of (\ref{metric00}). Eq. (\ref{EOM2}) leads to
\begin{equation}
F'_G= c r\,,
\end{equation}
c being a constant.
By dividing Eq.~(\ref{EOM1}) with respect to $2\mathrm{e}^{\alpha(r)} r^2$ and then
performing the derivative with respect to $r$, we obtain
\begin{equation}
c\left(\frac{4}{r^2}\left(\frac{d B(r)}{d r}\right)^2
+\frac{4 B(r)}{r^2}\frac{d^2 B(r)}{d r^2}-\frac{4(3B(r)-k)}{r^3}\frac{d B(r)}{d r}
\right)+\frac{\tilde Q^2}{r^5}=0\,.\label{equation} 
\end{equation}
In the topological case $k=0$ it is possible to find the general solution of such equation, namely
\begin{equation}
B(r)=\frac{1}{\sqrt{10}}\sqrt{C_1 -C_2 r^4-\frac{\tilde Q^2}{c r}}\,.\label{BBBB}
\end{equation}
The Gauss Bonnet invariant reads
\begin{equation}
G=-\frac{2}{5 c}\left(\frac{\tilde Q^2}{r^5}+6c C_2\right)\,,
\end{equation}
such that we can write $r$ as a function of $G$,
\begin{equation}
r=\left(\frac{2}{-5 c G-12 c C_2}\right)^{1/5}\tilde Q^{2/5}\,,
\end{equation}
and by using Eq. (\ref{EOM1}) we finally obtain that the model
\begin{equation}
F(G)=-\frac{1}{2}\left(\frac{-5cG}{2}-6c C_2\right)^{4/5}\tilde Q^{2/5}\,,
\end{equation}
exhibits the exact SSS Maxwell solution (\ref{metric00}) described by (\ref{BBBB}) for the topological case $k=0$. Note that $C_2$ is a fixed parameter of the model and $C_1$ is the free constant of the solution.

\section{Conclusions}
If some extended theory of gravity lies behind our universe,
the investigation of mathematical structure of this kind of theories becomes of great interest and the physics of black holes is a fashinating field of reasearch.
One should note that
in order to verify the consistence of a gravity theory in the weak-field limits with the standard solar-system tests of General Relativity, it is extremely important to know all the possible generalizations of the analogue of the Schwarzschild metric in the framework of such theory. In addition, the investigation of SSS metric Ansatz may be crucial also from the observational point of view, for example with regard to strong gravitational lensing, where some observable effects beyond those of General Relativity could be found and used to prove or disprove these theories.
In this paper, the lagrangian derivation of the Equations of Motion for studying topological SSS metrics in $\mathcal F (R, G)$-gravity has been presented. Due to a suitable form of the EOM, the study of the solutions results to be sensibly simplified. In particular, we found a new non trivial solution which may also describe the black holes in the topological case $k=1$. 
In this case, we
defined the BH mass with a quantity proportional to the constant of integration, which appears in the explicit solution, making use of the First Law of black hole thermodynamics and evaluating independently the entropy via Wald method and the Hawking temperature via quantum mechanical methods in curved space-times. We also showed that in principle may be possible to recover the First Law from the EOM, but, differently to the case of $F(R)$-gravity, the validity of this derivation is restricted to few (trivial) cases.
Finally, an exact solution for a specific Gauss-Bonnet modified gravity model in the presence of Maxwell field has been found.

\subsection*{Acknowledgements}
LS thanks Dr. Roberto Di Criscienzo and Prof. Guido Cognola for help and valuable suggestions and would like to appreciate the hospitality 
at Eurasian National University where the work was developed.

\end{document}